# Ferromagnetism in layered metallic $Fe_{1/4}TaS_2$ in the presence of conventional and Dirac carriers

Jin-Hua Wang（王金华）[1,2], Ya-Min Quan（全亚民）[1],

Da-Yong Liu（刘大勇）[1], Liang-Jian Zou（邹良剑）[1,2*]

[1]Key Laboratory of Materials Physics, Institute of Solid State Physics, Chinese Academy of Sciences, Hefei 230031, China

[2]Science Island Branch of Graduate School, University of Science and Technology of China, Hefei 230026, China

**Abstract** In this paper we present the microscopic origin of the ferromagnetism of $Fe_{0.25}TaS_2$ and its finite-temperature magnetic properties. We first obtain the band structures of $Fe_{0.25}TaS_2$ by the first-principles calculations and find that both conventional and Dirac carriers coexist in metallic $Fe_{0.25}TaS_2$. Accordingly, considering the spin-orbit coupling of Fe 3d ion, we derive an effective RKKY-type Hamiltonian between Fe spins in the presence of both the conventional parabolic-dispersion and the Dirac linear-dispersion carriers, which contains a Heisenberg-like, an Ising-like and an XY-like term. In addition, we obtain the ferromagnetic Curie temperature $T_c$ by using the cluster self-consistent field method. Our results could address not only the high ferromagnetic Curie temperature, but also the large magnetic anisotropy in $Fe_xTaS_2$.

**Introduction**

In the past decades great interest has focused on layered compounds known as *van der Waals* materials due to the weak bonding between layers[1-2], among which, transition metal dichalcogenides (TMDs) with structure $MX_2$ (M= transition metal and X=S, Se, Te) are typical hexagonal lattice of *van der Waals* materials. TMDs exhibit a wide spectrum of physical properties, including charge-density-wave $TaS_2$ and $TiSe_2$[3], insulator in $HfS_2$[4], semiconductors in $MoS_2$[5] and $WS_2$[6], semimetals in $WTe_2$[7] and $TcS_2$[8], metals in $NbS_2$[9] and $VSe_2$[10], superconductors in $NbSe_2$[11] and $2H-TaS_2$[12], and so on. The weak interlayer interaction in these *van der Waals* compounds allows the intercalation of charged ions or molecules between the $MX_2$ layers of the TMDs to significantly modify their physical properties: for example, Cu intercalation may drive charge-density-wave $TiSe_2$ into superconducting phase[13-14]. These TMDs have wide applications in spintronics, energy storage devices, catalysts, as well as optoelectronics[15-16].

Once intercalating 3*d*-transition-metal magnetic ions, long-range magnetic order may occur in $NbS_2$, $TiS_2$ and $TaS_2$[3,17]. Among them, $Fe_xTaS_2$ is a class of layered TMD compounds in which $Fe^{2+}$ ions are intercalated between the $TaS_2$ layers of $2H-TaS_2$. In 2006, Cava and Zandbergen et al. found that in $Fe_{1/3}TaS_2$ ferromagnetic order transition temperature is around 35 K, while the ferromagnetic Curie temperature of $Fe_{0.25}TaS_2$ compound is almost five times large, around 160 K.[18] Many studies have shown that the magnetic properties of $Fe_xTaS_2$ display complex behaviors with the amount of intercalant *x*[19-21]: it is a spin glass for *x*<0.2[22], a ferromagnet for 0.2<*x*<0.4[19], and an antiferromagnet for *x*>0.4[23]. In the ferromagnetic regime, the Curie temperature Tc strongly depends on Fe concentration: Tc increases with the increase of Fe content *x* for 0.2<*x*<0.26, while it decreases as *x* further increases[23]. Beyond that, many experimental results have shown that there are many interesting anomalies in ferromagnetic $Fe_xTaS_2$ with the insertion of Fe between the $TaS_2$ layers. On the one hand, magnetization, magnetoresistance, magnetic coercive force and effective magnetic moment increase with the increasing

Fe concentration; on the other hand, the magnetization of $Fe_xTaS_2$ sharply switches, and its extremely large magnetocrystalline anisotropy is comparable to that of rare-earth permanent magnets. These suggest that $Fe_xTaS_2$ is a potential candidate to substitute expensive rare-earth permanent magnet[23-27].

Such a potential application also stirs more theoretical interests, focusing on the microscopic origin of ferromagnetism and strongly magnetocrystalline anisotropy[25-30]. Zhu et al. showed that strong spin-orbit coupling of Fe $3d^6$ electronic configurations in the action of triangular crystal field is effective to enhance the magnetic anisotropy[28]; Cava et al. suggested the nature of ferromagnetic coupling of the system is the Ruderman-Kittle-Kasuya-Yosida (RKKY) interaction[25]. However, it is not clear why $T_c$ exhibits non-monotonic behavior on Fe concentration $x$, and why the $T_c$ of $Fe_{0.25}TaS_2$ can be as high as 160 K: The $T_c$ calculated by the first-principles LDA+DMFT method is about 10-20 K[43], which is very different from the experimental data. These questions appeals for a further theoretical investigation. In this Letter, we first present the band structures of $Fe_{0.25}TaS_2$, and find that both conventional parabolic and Dirac linear carriers coexist in the metallic ground state, and extract some electronic properties; from which we derive an effective Hamiltonian which includes the Heisenberg-type, Ising-type and XY-type terms between Fe spins; we further obtain the ferromagnetic Curie temperature $T_c$, and find that it is comparable with experimental data.

**The crystal and electronic structures of $Fe_{0.25}TaS_2$**

In the previous works, Cava et al.[18] and Ong et al.[24] found that $Fe^{2+}$ ions are periodically distributed in the double cells of parent compound $TaS_2$, rather than randomly scattered between the $TaS_2$ layers. $Fe^{2+}$ ions form a triangular layer by crystallizing in a 2×2 superlattice within the interstitial space between $TaS_2$ layers, which can be obviously seen from Fig.1. In the present compound the Fe spins constitute a two-diemnsional triangular lattice. The detailed crystal structure parameters is listed in Table I and crystal structure of $Fe_{0.25}TaS_2$ is shown in Fig. 1, respectively.

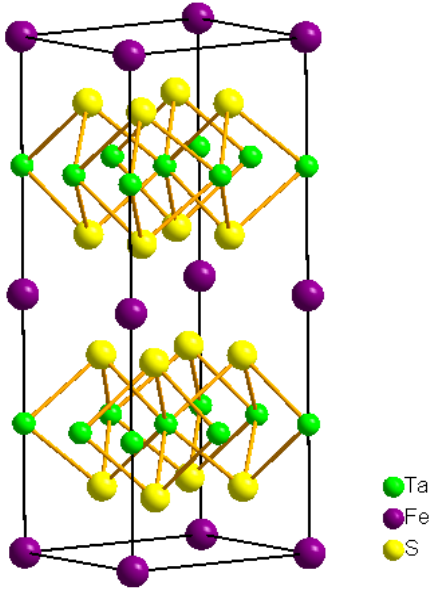

**FIG. 1**: The crystal structure of $Fe_{0.25}TaS_2$. Yellow, green, and violet balls represent S, Ta, and Fe atoms, respectively. Intercalated Fe ions form an ordered superlattice, giving rise to superstructures with $a'=2a_0$, where $a_0$ is the basic hexagonal lattice parameter of the $TaS_2$ array.

**TABLE I**. Crystal structure parameters of $Fe_{0.25}TaS_2$ with space group (no.194) P63/mmc, a=6.6141(15)Å, c=12.154(3) Å. Z=2, $R_F$=0.037.[18]

| Atom | x | Y | z | $B_{iso}$ |
|---|---|---|---|---|
| Ta（1） | 0.49507(6) | 0.50493 | 0.75 | 0.520(16) |
| Ta（2） | 0 | 0 | 0.25 | 0.511(19) |
| Fe | 0 | 0 | 0 | 1.07(8) |
| S（1） | 2/3 | 1/3 | 0.1191(4) | 0.64(8) |
| S（2） | 0.83177(25) | 0.16823 | 0.62233(2) | 0.57(6) |

To elucidate the microscopic origin of the strong ferromagnetism in $Fe_xTaS_2$, one should understand its essential band structures. In obtaining the electronic structures of $Fe_{0.25}TaS_2$, we use the generalized gradient approximation (GGA) method within the density functional theory. The projected band structures are shown in Fig.2, where the thick lines denote the Fe 3d orbits. We find several major characters in the energy band of $Fe_{0.25}TaS_2$ in Fig.2: (1) around $E_F$ there exist almost flat bands coming from Fe 3d orbits, and the flat band dominantly comes from $d_{x^2-y^2/xy}$ orbits and a part of $d_{xz/yz}$ orbits, it contributes to the local spin of Fe ion; (2) there are two kinds of conducting carriers near the Fermi surface, one is the conventional carrier with the parabolic energy dispersion, the other is the unusual Dirac carrier with the linear dispersion, as indicated by the red circles in Fig.2; (3) the center of these flat bands crosses with conduction bands, showing that the conducting carriers considerably hybridize with local spins, as seen in Fig. 2. These characters point to the RKKY-type indirect magnetic coupling of the local spins, in accordance with the suggestion by Cava et al. and Cheong et al.. Since the distance between Fe spins is rather far away, this fact rules out the possibility of direct exchange interaction between spins.

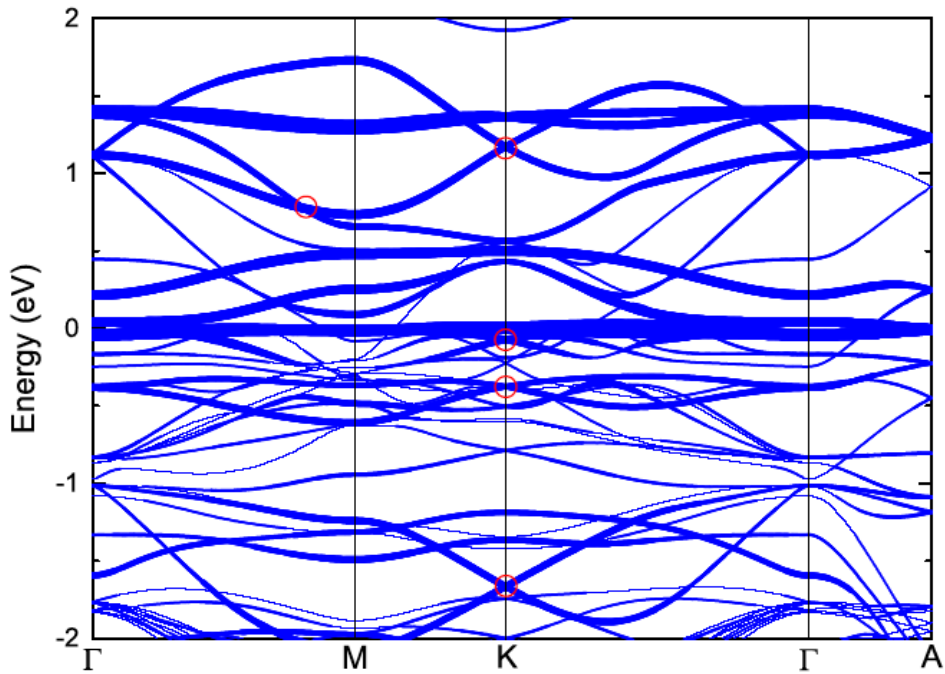

**FIG.2:** The band structures of ferromagnetic metallic phase in $Fe_{0.25}TaS_2$ in the absence of spin-orbital coupling. The heavy lines indicate the weights of Fe 3d orbitals. The red circles indicate the Dirac points.

**The RKKY interaction and the effective model**

Our numerical study shows that in $Fe_{0.25}TaS_2$, the distance between Fe spins is rather far away, the RKKY interaction strength of the nearest-neighbor Fe spins contributing from the parabolic energy band is only about one-fifth or less of that from the linear energy band, suggesting rather weak and negligible. Based on this fact, we only consider the contribution of the Dirac-type linear energy band on the RKKY interaction in what fellows. We pay particular attention to the contribution of Dirac carriers to the RKKY interaction in the presence of the spin-orbit coupling (SOC). In order to investigate the strength of the RKKY interaction in the presence of both the Dirac-type linear energy band and the conventional parabolic energy band, we take the nearest-neighbour distance of Fe spins $R_{12}$=6.6141 Å, the Fermi velocity $\hbar v_F$=0.3774 eV·Å, the exchange coupling between conducting carriers and local spins $J$=0.1 eV and the Fermi energy $\varepsilon_F$=0.2 eV.

We now derive the RKKY interaction between two magnetic impurities located in linear energy band system in the presence of the SOC. Using retarded Green's functions in real space, we acquire asymptotic expressions for the interaction. We find that the spin-orbit coupling has significant influences on the RKKY interaction, resulting in a Heisenberg-type, an Ising-type, and an additional XY-type term. We consider the magnetic impurities $S_i$ (i=1,2) located at $R_i$ mediated by Dirac electrons on a plane of 3D materials. We start from the following Hamiltonian,

$$H = H_0 + H_i^{int} \quad (1)$$

where $H_0$ contains the kinetic energy of the linear-dispersion carriers and spin-orbit coupling

$$H_0 = \chi \hbar v_F \vec{k} \cdot \vec{\sigma} + \alpha \vec{n} \cdot (\vec{k} \times \vec{\sigma}), \chi = \pm 1 \quad (2)$$

and $H_i^{int}$ denotes the s-d interaction between the magnetic impurities and the conducting carriers

$$H_i^{int} = J\vec{S}_i \cdot \vec{\sigma} \delta(\vec{r} - \vec{R}_i) \quad (3)$$

Here χ is the chirality of the Dirac carriers, $k$ is its in-plane momentum; $\vec{\sigma} = (\sigma_x, \sigma_y, \sigma_z)$ is the Pauli matrix denoting real spin of the Dirac carrier; $v_F$ is the Fermi velocity of the linear-dispersion electrons, which is about $\hbar v_F$=0.3774 eV·Å for $Fe_{0.25}TaS_2$; α represents the strength of the SOC, which is around 0.06×2√3 eV·Å; J denotes the strength of the s-d exchange interaction, which is around 0.1 eV in our work. We take $\vec{n} = \vec{z}$ as the unit vector along the normal direction of the surface.

We use retarded Green's functions to derive the RKKY interaction between Fe in $Fe_{0.25}TaS_2$. Here we rewrite the wave vector **k** according to the direction along the real coordinate **R** and the direction vertical to **R**: $\vec{k} = \vec{k}_{\parallel} + \vec{k}_{\perp} = (\vec{k} \cdot \vec{e}_R)\vec{e}_R + (\vec{e}_R \times \vec{k}) \cdot \vec{e}_R$. Then, the Green's function in real space can be obtained as

$$G_{\chi}(\pm \vec{R}; \varepsilon) = \sigma_0 G_0(R; \varepsilon) \pm [\chi G_1(R; \varepsilon)\vec{\sigma} \cdot \vec{e}_R + G_2(R, \varepsilon)(\vec{n} \times \vec{e}_R) \cdot \vec{\sigma}] \quad (4)$$

where

$$G_0(R,\varepsilon) = -\frac{\varepsilon}{4\pi R(\hbar^2 v_F^2 + \alpha^2)} e^{i\frac{\varepsilon R}{\sqrt{\hbar^2 v_F^2 + \alpha^2}}} \quad (5)$$

$$G_1(R,\varepsilon) = -\frac{\hbar v_F \varepsilon R + i\hbar v_F \sqrt{\hbar^2 v_F^2 + \alpha^2}}{4\pi R^2 \sqrt{(\hbar^2 v_F^2 + \alpha^2)^3}} e^{i\frac{\varepsilon R}{\sqrt{\hbar^2 v_F^2 + \alpha^2}}} \quad (6)$$

$$G_2(R,\varepsilon) = -\frac{\alpha \varepsilon R + i\alpha \sqrt{\hbar^2 v_F^2 + \alpha^2}}{4\pi R^2 \sqrt{(\hbar^2 v_F^2 + \alpha^2)^3}} e^{i\frac{\varepsilon R}{\sqrt{\hbar^2 v_F^2 + \alpha^2}}} \quad (7)$$

and $\vec{e}_R \equiv \frac{\vec{R}}{R}$. In the loop approximation[31-32] we find the RKKY interaction between two magnetic impurities in the form of

$$\begin{aligned}H_{1,2}^{RKKY} &= -\frac{2}{\pi} \text{Im} \int_{-\infty}^{\varepsilon_F} d\varepsilon Tr[H_1^{int} G_{\chi}(\vec{R}; \varepsilon + i0^+) H_2^{int} G_{\chi}(-\vec{R}; \varepsilon + i0^+) \\ &= -\frac{J^2}{\pi} \text{Im} \int_{-\infty}^{\varepsilon_F} d\varepsilon Tr[(\vec{S}_1 \cdot \vec{\sigma}) G_{\chi}(\vec{R}_{12}; \varepsilon + i0^+)(\vec{S}_2 \cdot \vec{\sigma}) G_{\chi}(-\vec{R}_{12}; \varepsilon + i0^+)\end{aligned} \quad (8)$$

Where $\varepsilon_F$ is the Fermi energy, $\vec{R}_{12} = \vec{R}_1 - \vec{R}_2$, and $Tr$ means a trace over the spin degrees of freedom of conduction electrons. Then the RKKY interaction can be written as

$$H_{1,2}^{RKKY} = F_1(\xi_F)\vec{S}_1 \cdot \vec{S}_2 + F_2(\xi_F)S_1^z S_2^z + F_3(\xi_F)(\vec{S}_1 \times \vec{S}_2)_z, \quad (9)$$

where the range functions are

$$F_1(\xi_F) = -\frac{9\pi \varepsilon_F^5 J^2}{8k_F^6 \xi_F^7 (\hbar^2 v_F^2 + \alpha^2)^4} \left[\frac{\xi_F^2(\hbar^2 v_F^2 + 11\alpha^2) - \xi_F^4(2\hbar^2 v_F^2 + 6\alpha^2)}{2}\cos(2\xi_F) + \xi_F^3(\hbar^2 v_F^2 + 7\alpha^2)\sin(2\xi_F)\right] \quad (10)$$

$$F_2(\xi_F) = -\frac{9\pi \varepsilon_F^5 J^2 \alpha^2}{4k_F^6 \xi_F^5 (\hbar^2 v_F^2 + \alpha^2)^4}[(5 - 2\xi_F^2)\cos(2\xi_F) + 6\xi_F \sin(2\xi_F)] \quad (11)$$

$$F_3(\xi_F) = -\frac{9\pi\varepsilon_F^5 J^2 \alpha}{2k_F^6 \xi_F^5 (\hbar^2 v_F^2 + \alpha^2)^{7/2}}[(\xi_F^2 - 1)\sin(2\xi_F) + 2\xi_F \cos(2\xi_F)] \tag{12}$$

respectively, where $\xi_F = \frac{\varepsilon_F R_{12}}{\sqrt{\hbar^2 v_F^2 + \alpha^2}}$. It is obvious that the RKKY interaction consists of three terms: the Heisenberg-type term, the Ising-type term and the XY-type term, each displays different range function. Among the anisotropic interactions in Eqn.(9), the Dirac carriers contribute the Ising-type term, and the spin-orbit coupling results in the XY-type term.

Considering the Fe $3d^6$ electron configuration, the Fe spins form a spin-2 quasi-two-dimensional ferromagnet on a triangular lattice in $Fe_{0.25}TaS_2$. The effective model Hamiltonian is thus expressed as:

$$\begin{aligned}H = &J_1 \sum_{<ij>}[(1-\Delta_1)(S_i^x S_j^x + S_i^y S_j^y) + S_i^z S_j^z] + J_2 \sum_{<ij>}[(1-\Delta_2)(S_i^x S_j^x + S_i^y S_j^y) + S_i^z S_j^z] \\ &+ J_{xy}^1 \sum_{<ij>}(S_i^x S_j^y + S_i^y S_j^x) + J_{xy}^2 \sum_{<ij>}(S_i^x S_j^y + S_i^y S_j^x) + D\sum_i (S_i^z)^2\end{aligned} \tag{13}$$

where the summations run over the first- and the second-nearest-neighbor pairs, $J_i(i=1,2)$ denote the strengths of the first- and the second-nearest-neighbor Heisenberg-type and the XY-type interactions, respectively; $\Delta_i = \frac{J_i^z}{J_i}, (i=1,2)$ represent the strength ratios of the Ising-type interaction with respect to the Heisenberg term; and $D$ represents the magnetic anisotropy. We find that due to the fast power-law decay, the first two nearest-neighbor magnetic interactions are considerable, the other long-range interactions are small and negligible.

**The finite-temperature magnetic properties**

Having the effective spin-spin interactions, one could discuss the finite-temperature magnetic properties of $Fe_xTaS_2$. Substituting $J$=0.1 eV, $\hbar v_F$=0.3774 eV·Å, $\alpha = 0.06 \times 2\sqrt{3}$ eV·Å, $\varepsilon_F$=0.2 eV, and $R_{12}$=6.6141 Å into the range functions (10), (11) and (12), we obtain the values of the parameters, $J_1, J_2, \Delta_1, \Delta_2, J_{xy}^1$ and $J_{xy}^2$, which are listed in Table II

**TABLE II.** Magnitudes of the exchange coupling parameter in the effective Hamiltonian in Eq.(13)

| parameters $\varepsilon_F$/eV | $J_1$/meV | $J_2$/meV | $\Delta_1$ | $\Delta_2$ | $J_{xy}^1$/meV | $J_{xy}^2$/meV |
|---|---|---|---|---|---|---|
| 0.20 | -3.80 | -0.05 | -0.5800 | -2.000 | 2.80 | 1.00 |

Starting from the effective model in Eq.(13), we calculate the magnetic moment as a function of temperature on triangular lattices using the self-consistent cluster method[33] with the finite temperature Lanczos method as the solver[34-39]. Considering that the system is ferromagnetically ordered, all the lattices only consider meeting the periodic boundary conditions, as shown in Fig. 3. We perform the numerical calculation to obtain the T-dependent magnetization of the triangular ferromagnet with the cluster sizes N=9 and 12, respectively, so as to obtain the critical temperature of the ferromagnet-paramagnet phase transition. In our calculation, since the computer memory and costing time for the spin clusters with S=2 are rather huge, instead, we use spin-1/2 ferromagnetic clusters as substitutions. After obtaining $T_C$ for spin-1/2 cluster, we scale by multiplying 4S(S+1)/3 to obtain the $T_C$ for spin-2 clusters.

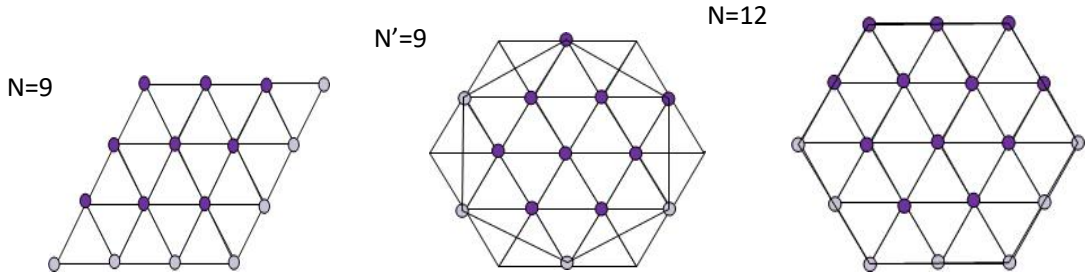

**FIG.3**: The cluster sizes studies in this work. The deep color symbols denote the triangular lattice sites, while the light color symbols are redundant and related to filled ones via cell translation vectors.

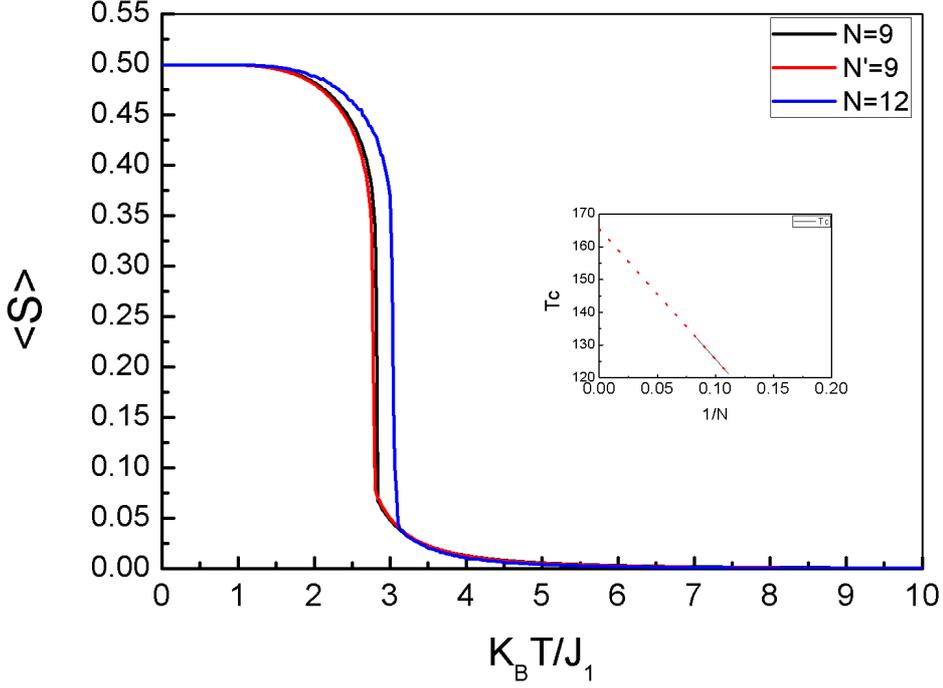

**FIG.4**: The temperature dependence of the local spins at different cluster sizes: black and red for N=9, blue for N=12). The temperature is measured in the unit of $|J_1|$.

As shown in Fig.4, the magnetization gradually decreases with the lifting temperature, and critically reduces to extremely small. The presence of anisotropic exchange interaction leads to the ferromagnetic-paramagnetic phase transition first order. We define the ferromagnetic Curie point $T_c$ at the sharply decreasing temperature of the magnetization <S>. Naturally, from the finite-size scaling in the inset of Fig.4, one could extrapolate that the Curie temperature occurs at $T_c \approx 165$ K in an infinite system. This result is more consistent with the experimental results, in comparison with the LDA+DMFT result[43]. This arises from fact that the LDA+DMFT method itself could not take into the long-range interaction like the RKKY coupling. These results manifest that the presence of Dirac linear-dispersion carriers and the SOC is fundamental to the ferromagnetism and its high Curie temperature of $Fe_{0.25}TaS_2$.

Thus, one could understand why the Curie temperature of $Fe_{0.25}TaS_2$ is so particular high in comparison with other Fe doping concentration, since there exists hexagonal symmetry in the Fe layers, leading to the presence of Dirac carriers. These unusual carriers mediate the strong ferromagnetic interaction between Fe spins, and contribute magnetic anisotropy to stabilize the magnetic order. It is also expectable that once x deviates from 1/4, the ferromagnetic temperature of $Fe_xTaS_2$ considerably decreases, as found in experiments [18, 29].

We also notice that in $Fe_{0.25}TaS_2$ the Dirac state comes from the six-fold high symmetry of triangular lattice of Fe spins and is protected by the crystal structure. The presence of long-range ferromagnetic order and the breaking of the time-reversal symmetry could not drive the Dirac state into the Weyl state in the present compound. This situation is different from the Dirac state originated from the SOC in other topological materials. Meanwhile our first-principles calculations show that after considering the SOC, the Dirac states will open a small gap. However, this does not affect our results since as a medium, the Dirac states are integrated out. It leaves the an anisotropic RKKY interaction in Eqn.(9).

In conclusion, we show that metallic $Fe_{0.25}TaS_2$ is an anisotropic RKKY ferromagnet with the Dirac linear-dispersion carriers. The Dirac carriers participating in the RKKY interaction contribute to the Ising-type spin coupling, and the relativistic spin-orbit coupling leads to the anisotropic XY-type terms in the RKKY interaction. These interactions contribute to the high Curie temperature in ferromagnetic $Fe_{0.25}TaS_2$. We believe that the present theory is also applicable to many other quasi-two-dimensional magnetic materials with both the Dirac carriers and strong spin-orbit coupling.

**Acknowledgements:**
This work is supported by the National Natural Science Foundation of China under Grant Nos. 11774354, and 11534010 and 11574315. Numerical calculations were performed at the Center for Computational Science of CASHIPS and Tianhe II of CSRC.